\documentclass[symmetry,article,accept,pdLaftex,moreauthors]{Definitions/mdpi} 

\firstpage{1} 
\pubvolume{1}
\issuenum{1}
\articlenumber{0}
\pubyear{2026}
\copyrightyear{2026}
\externaleditor{Firstname Lastname}
\datereceived{7 December 2025} 
\daterevised{16 January 2026} 
\dateaccepted{19 January 2026} 
\datepublished{ } 
\hreflink{https://doi.org/} 

\Title{The Hubble Tension in Light of the Symmetry of Scale~Invariance}

\newcommand{\kmsmpc}{km s$^{-1}$ Mpc$^{-1}$}
\newcommand{\om}{$\Omega_{\mathrm{m}}$}

\Author{Frédéric Courbin $^{1,2,3}$\orcidA{} and  André Maeder  $^{4,}$*  \orcidB{}} 

\address{%
$^{1}$ \quad {ICC-UB Institut de Ci\`encies del Cosmos, Universitat de Barcelona, Mart\'i Franqu\`es, 1, 08028 Barcelona, Spain; frederic.courbin@icc.ub.edu} \\
$^{2}$ \quad Institut Català de Recerca i Estudis Avançats (ICREA), 
 Pg. Llu\'is Companys 23, 08010 Barcelona,  Spain \\
$^{3}$ \quad
Institut d’Estudis Espacials de Catalunya (IEEC), Edifici RDIT, Campus UPC,  Castelldefels, \linebreak 08860 Barcelona, Spain \\
$^{4}$ \quad Department of Astronomy, University of Geneva, Chemin Pegasi 51, CH-1290 {Versoix,}  Switzerland
}

 \corres{Correspondence: frederic.courbin@icc.ub.edu, andre.maeder@unige.ch}

\abstract{{When the expansion rate of the Universe at recombination is used to infer the present expansion rate $H_0$, 
the value derived in the $\Lambda$CDM model, $H_0=67.4$ \kmsmpc, is about in 6$\, \sigma$ tension with the value measured locally, $H_0=74$ \kmsmpc. In this work, we consider instead the expansion history in the context of the symmetry of scale-invariant vacuum (SIV model). We first perform two major cosmological tests: the Hubble diagram for type-Ia supernovae and the fundamental relation between $H_0$, the age of the Universe, and the total density of matter, \om. This allows us to constrain \om\ in SIV, with both tests giving the  best agreement for \om   ~= 0.20. We then study the physical connections of the dynamical and thermal states of the Universe at recombination with the present Hubble constant, $H_0$, and the present temperature, $T$, in the $\Lambda$CDM and SIV contexts. We find that, in SIV, the properties at recombination may be conveyed to the present ones ($T=2.726$ and $H_0$ at $z=0$) without any tension,  indicating $H_0=74$ \kmsmpc\, in spite of the anchoring on the CMB. This is due to the slightly different  expansion  and temperature histories of the two cosmological models. Importantly, this happens to occur for \om ~= 0.20, as constrained in SIV with supernovae and cosmic age. This suggests that the Hubble tension currently found between $H_0$ values in the early and late Universe may simply be the result of $\Lambda CDM$ ignoring the small but still measurable effects of scale invariance.}}


\begin{document}


\section{Introduction} 
\label{intro}

Hubble tension, the difference between the value of the   Hubble parameter  \linebreak $H_0=74$ \kmsmpc determined in the late universe \citep{Breuval2024} and the value of \linebreak $ H_0= 67.4$  \kmsmpc, determined from the CMB fluctuations in the early universe~\citep{Planck21}, has progressively gained a high level of significance. This has been strengthened even further with recent CMB observations from SPT-3G in E-mode polarization and high spatial frequencies, giving $H_0 = 66.66\pm0.60$ \kmsmpc ~\citep{Campuis2025}. This is in 6.2 $\sigma$ tension with local~measurements.

The tension now seems clearly established and acknowledged as the signature of a severe problem in cosmology and physics rather than an observational fluke. Thorough reviews by \citet{Verde19}, \citet{KamionRiess23}, and \citet{diValentino2025} examine the main observational constraints, the possibly discrepant points in the physics of the early and late Universe, as well as possible suggestions of new physical effects to be considered in the future. Considering the many methods to measure $H_0$, \citet{Verde19} emphasizes that, ``This shows that the discrepancy does not appear to be dependent on the use of any one method, team, or source''. Also, these authors point out that the discrepancy seems to concern   the early physics and the possible changes affecting the expansion phase preceding the end of the radiative phase.

On the observational side, the validity of the determination of the Hubble constant  has been further supported by the Large Magellanic Cloud Cepheid Standards with great accuracy \citep{RiessCaser19}. Potential errors in the Cepheid calibration or about  the possibility of local  density voids affecting measurements of the local $H_0$ have been disregarded by new Gaia DR3 parallaxes \citep{RiessCaser18, RiessCaser21}. Moreover, JWST observations of a sample of more than 1000~Cepheids as an  anchor of the distance ladder confirm within 0.01 magnitude the distance scale previously adopted by SH0ES \citep{RiessAnand24}.  New observations  \citep{Scolnic25} with the Dark Energy Spectroscopic Instrument (DESI)  of the distance to the Coma Cluster support  a rather high value $H_0 = 76.5 \pm 2.2$ \kmsmpc, even more differing from the  CMB Planck value. Very recently, 
a study with detailed error audit of the measurements even reported a 7.1$~\sigma$ tension \cite{H0DN}. All these observations call for cosmological developments beyond\linebreak   the $\Lambda$CDM model.

Numerous original and inventive cosmological models have been studied regarding the physics of the end of the radiative phase, such as the ``Early Dark Energy (EDE)'' model, which proposes a transitory background peak of some unknown dark energy phase being active just before recombination and essentially insignificant in  earlier and later phases. Further theoretical  models are proposed with evolving fluid densities playing a significant role at some cosmological stages and also new models where either transitory, oscillating, or rolling new scalar field are envisaged in the radiative phase \citep{KamionRiess23}. However, alternatives to the current $\Lambda$CDM paradigm should at least respect two crucial rules: 

\begin{itemize}

\item \emph{``A mathematical principle, logical and simple, (must) totally or nearly determine the equations (of gravitation)"},  according to Einstein \citep{Einstein49}.

\item \emph{"The new model should not do worse than standard $\Lambda$CDM in describing all other cosmological observations"}, as emphasized by \citet{Verde19}.

\end{itemize}

In the present work, we explore scale-invariant models (``Scale Invariant Vacuum''; SIV models) \citep{Maeder17a,MaedGueor23}. SIV clearly obeys the first requirement, as the whole theory flows naturally when considering the question: In addition to Galilean invariance, to Lorentz invariance and to general covariance,
is there still another invariance to consider in the gravitation theory? In the empty space, general relativity is also scale-invariant.
However, the presence of mass generally suppresses scale invariance \citep{Feynman63}. How small has to be the density not to destroy  scale invariance? Cosmological models based on the scale-invariant field equation by \citet{Dirac73} indicate a drastic reduction of the effects of scale invariance 
as soon as  even some negligible amount of matter is present \citep{Maeder17a}. However, although very much reduced, 
these effects totally vanish only at the
critical density  $\rho_{\mathrm{c}= }\frac{3 H^2_0}{8\pi G}$ ($\sim 10^{-29}$ g cm$^{-3}$) and above. This limit is consistent  
 with considerations about causal connection in the Universe \citep{MaedGueor21a}.

The second of the above two requirements has been progressively verified by confronting SIV with key observational tests. Noticeable examples include the growth of density fluctuations \citep{MaedGueor19}, the dynamics of galaxies \citep{Maeder23} and of galaxy clusters \citep{MaedCourb24}. In addition, the departures from classical mechanics measured in the motion of very wide binary stars \citep{Hernandez23,Chae23} were found consistent with the predictions of SIV \citep{MaedCourb24}. In the latter work, detailed tests were also performed with strong gravitational lensing, showing that scale-invariant effects were able to reconcile the lensing and stellar mass of lens galaxies, both at low redshift in the SLACS sample and at redshift z$\sim$2 in the JWST-ER1 system. 

In the present work, we explore the expansion history of the Universe, $H(z)$ as described in SIV and show that the Hubble tension is nicely explained even when anchoring $H(z)$ on the CMB. Section~\ref{SIV} presents the basic analytical properties of the SIV models, and  some relevant  model properties of the early Universe. Section \ref{sec:obs} discusses the SNIa observations
and the relation between $H_0$, \om, and the age of the Universe, in light of SIV. Section~\ref{tension} explores the origin of the Hubble tension and Section~\ref{conclusion} presents our conclusions.

\section{The Scale-Invariant Models}  
\label{SIV}

Although the full SIV theory has already been published \citep{Maeder17a,MaedGueor23} along with all the details of equation derivations, we briefly summarize the main points relevant to Hubble tension. 

\subsection{Short  Basics} 
\label{base}

The scale-invariant theory assumes general covariance like general relativity; in addition, it also assumes scale covariance to a form 
\begin{equation}
ds' \, = \, \lambda \, ds\,.
\end{equation}
$ds'$ is the line element in the space-time of GR, while $ds$ is the scale-transformed line element in a new space-time, endowed with the possibility of scale transformation by a factor of $\lambda$. The notion of cotensor  was introduced by Dirac \citep{Dirac73}: mathematical bodies  possibly subject to general covariance  and to scale covariance of the form $Y' =\lambda^n  Y$. For $n=0$ one speaks of scale invariance and of  intensor, invector or inscalar. A cotensoral expression of the scale-invariant field equation has been developed by Dirac \citep{Dirac73} and Canuto et al. \citep{Canuto77}, it was further confirmed by an action principle \citep{MaedGueor23}. In this context, it is worth to emphasize some remarks by   \citet{Einstein49}  in his  Autobiographical Notes of 1949: \emph{``…the existence of rigid standard rulers is an assumption suggested by our approximate experience, assumption which is arbitrary in its principle''}.  

A gauging condition is necessary to constrain the scale factor. \citet{Dirac73} and \citet{Canuto77} adopted the so-called large number hypothesis. It was preferred to adopt as a gauge condition that ``the macroscopic empty space must be scale invariant'' \citep{Maeder17a}, which implies a gauge factor $\lambda \sim 1/t$. Moreover, this gauging condition is also demanded by the above mentioned action principle \citep{MaedGueor23}.

Cosmological equations have been obtained with a geometry described by the FLWR metric and with the gauging assumption that the macroscopic empty space is scale-invariant~\citep{Maeder17a},
\begin{equation}
\frac{8 \, \pi G \varrho }{3} = \frac{k}{a^2}+\frac{\dot{a}^2}{a^2}+ 2 \,\frac{\dot{a} \dot{\lambda}}{a \lambda} \, ,
\label{AE1}
\end{equation} 
\begin{equation}
-8 \, \pi G p  = \frac{k}{a^2}+ 2 \frac{\ddot{a}}{a}+\frac{\dot{a}^2}{a^2}
+ 4 \frac{\dot{a} \dot{\lambda}}{a \lambda}  \, .
\label{A2}
\end{equation}
%

The combination of these two equations  leads to
\begin{equation}
-\frac{4 \, \pi G}{3} \, (3p +\varrho)  =  \frac{\ddot{a}}{a} + \frac{\dot{a} \dot{\lambda}}{a \lambda}  \, .
\label{E3}
\end{equation}

 Each  equation contains only one additional term with respect to Friedmann's.
The last equation shows that this additional term  produces an acceleration proportional to the velocity.  Of importance for the Hubble expansion, we see that the above system of scale-invariant 
cosmological equations  naturally contains  an acceleration term, which results from the (small) variations in the scale factor $\lambda$
with time. In the context of the scale-invariant models, it is therefore not necessary to call for  different possible forms of dark energy.
\textls[-25]{An additional acceleration is also present in the weak field approximation, thus leading to a modified Newton equation \citep{MBouvier79} (see also~\citet{Maeder23, MaedGueor23}).} 
Such an  acceleration has also been found to favor the growth of density fluctuations
and thus the early formation of galaxies at very high redshifts~\citep{MaedGueor19}. 
The above gauging condition based on the properties of the empty space also implies that the scale factor $\lambda(t)$  is independent of the matter content.

A specific conservation law for SIV results  from the above cosmological equations,
\begin{equation}
\varrho \, a^{3(w+1)}  \,  \lambda ^{(3w+1)} \,= const,
\label{3w}
\end{equation}  
with $p= w \varrho \, c^2$ (c=1 in the three above cosmological equations).  This conservation law, together with the condition of the scale invariance of the energy-momentum $T_{\mu \nu}$, imposes  that the potential $\Phi = GM/r$ is a conserved quantity rather than the mass, $M$, which varies in $\lambda^{-1} = t$ \citep{Maeder17a,MaedGueor23}. Thus, as in special relativity, the mass is not a conserved quantity. However,  in current models, the amplitude of these variations over the age of the Universe is very limited.
\color{black}

The additional effects resulting from scale invariance would be  quite large if there were no matter in the Universe ($\Omega_{\mathrm{m}}=0$), with $\Omega_{\mathrm{m}}=\varrho/\varrho_{\mathrm{c}}$ and $\varrho_c= \frac{3H^2_0}{8 \pi G}$. However, a
noticeable property of these equations is that the effects of scale invariance are drastically reduced as soon as  some matter is  considered  in the models (see {Figure~3}  in \citep{Maeder17a}).  The effects  completely vanish for  models with $\Omega_{\mathrm{m}}=1$ (identical to the EdS model) and above. Moreover, in  models with $0< \Omega_{\mathrm{m}} < 1$, the scale-invariant effects in current weak-field dynamics behave like $ \sim  \sqrt{\varrho_c/\varrho}$,  implying a further drastic reduction in scale-invariant effects with the increasing density of the  systems considered \citep{Maeder17c}. Remarkably, the above results are  in agreement with the statements by \citet{Feynman63} about the occurrence of scale invariance in empty space, and with its rapid suppression in the presence of matter. 

The solutions of the above equations for $k=0$ and $\pm 1$ are close to those of the $\Lambda$CDM models when \om $ > 0.3$
  according to the numerical models \citep{Maeder17a}. For matter-dominated models with  $k=0$,  there is a simple  analytical solution   \citep{Jesus18},
\begin{equation}
a(t) \, = \, \left[\frac{t^3 -\Omega_{\mathrm{m}}}{1 - \Omega_{\mathrm{m}}} \right]^{2/3}\, , \quad \mathrm{with} \; \;
t_{\mathrm{in}}= \Omega^{1/3}_{\mathrm{m}},
\label{Jesus}
\end{equation}
where $t_{\mathrm{in}}$ is the initial time for $a(t_{\mathrm{in}})=0$. The solution for  models in the radiative phase of the Universe has been studied in
\citep{Maeder19}. For the present time  $t_0=1$, one also has $a(t_0)=1$.  It is somehow surprising that the addition of one more invariance in the theory of gravitation leads to simpler equations  than in the $\Lambda$CDM models (cf. Appendix~\ref{LCDM}).

{A word of caution is necessary about \om. Although its definition in SIV is similar to that in $\Lambda$CDM, it does not mean that the numerical values of the SIV density, temperature, and other properties are identical to those of the $\Lambda$CDM  models. The reason is that \om\, indirectly influences, through several equations, other quantities such as $H_0$ and the critical density $\varrho_{\mathrm{c}}$. It also influences the age of the Universe, $\tau_0$, the times of various important events, such as recombination, and eventually all model parameters and properties. For example, if we consider the cases of \om ~= 0.3 and \om ~= 0.2 in the $\Lambda$CDM models, the same values of the critical density, $\varrho_{\mathrm{c}}$, occur for \om ~= 0.28 and \om ~= 0.17 in the SIV models.  Thus, there is no exact numerical correspondence of all parameters between models of the same \om\ value in $\Lambda$CDM and SIV, particularly for \om ~$\leq 0.30$. This is why when we refer to \om\ we often speak of specific SIV \om\ values. The cosmological models being different, all their variables  evidently have (often slightly) different numerical values.}

The derivative of (\ref{Jesus}) leads to the Hubble expansion rate in the $t$ scale,
\begin{equation}
H(t)=  \frac{2 \, t^2}{t^3-\Omega_{\mathrm{m}}}, \quad \mathrm{thus} \; \; H(t_0)= \frac{2}{1- \Omega_{\mathrm{m}}}.
\label{h}
\end{equation}

The term $\Omega_{\mathrm{m}}$ represents the value at the present time $t_0=1$, alike in the $\Lambda$CDM. 
At present,  we have for $k=0$ \citep{Maeder17a},
\begin{equation}
\Omega_{\mathrm{m}}+ \Omega_{\lambda} = 1, \quad \mathrm{with} \quad  \Omega_{\lambda}= 1 - \frac{2}{H(t) \times t}.
\end{equation}

The dependence in $\Omega_\mathrm{m}^{1/3}$  of $ t_{\mathrm{in}}$ (Eq.~\ref{Jesus}) produces a rapid increase in $ t_{\mathrm{in}}$   for low $\Omega_{\mathrm{m}}$. For $\Omega_{\mathrm{m}}=0, 0.01, 0.1, 0.3, 0.5$, the values of $ t_{\mathrm{in}}$ are 0, 0.215, 0.464, 0.669, and 0.794, respectively. This strongly reduces the possible range of $\lambda(t)= t_0/t$, which varies between $1/t_{\mathrm{in}}$ and $1/t_0=1$  for increasing $\Omega_{\mathrm{m}}$ 
(cf. {Figure~3} in \citep{Maeder17a}). For  $\Omega_{\mathrm{m}} \geq 1$, there are no possible scale-invariant models, consistent with causality relations in an expanding Universe \citep{MaedGueor21a}.

The SIV models,  $a(t)$ vs. $t$,  are relatively close to the $\Lambda$CDM models for \om ~$> 0.3$  up to \om ~= 1, 
where they become identical  to the EdS model. For smaller \om,
departures become larger and larger with decreasing \om\ values.

\subsection{Evolution of Scale-Invariant Properties with Redshift $z$}
%
By expressing the redshift $z+1=a(t_0)/a(t)$,  one may obtain the dimensionless time $t$ as a function of $z$,
\begin{equation}
t^3=\Omega_{\mathrm{m}}+ (1-\Omega_{\mathrm{m}})  a^{3/2},  \; \mathrm{thus} \; 
t  =  \left[ \Omega_{\mathrm{m}}+ (1- \Omega_{\mathrm{m}})(1+z)^{-3/2} \right]^{1/3}.
\label{tz}
\end{equation}

To transfer time $t$ to the current time $\tau$  in seconds or years, we use expressions (\ref{T2}),
 which give 
\begin{equation}
\tau = \tau_0 \,  \frac{\left[\Omega_{\mathrm{m}}+(1-\Omega_{\mathrm{m}})(1+z)^{-3/2}\right]^{1/3} -\Omega^{1/3}_{\mathrm{m}}}
{1-\Omega_{\mathrm{m}}^{1/3}      }.
\label{taz}
\end{equation}

The current $H$ (in the $\tau$-scale) is
\begin{equation}
H(\tau)= \frac{da}{dt} \frac{dt}{d\tau} \frac{1}{a}= H(t) \frac{1-\Omega^{1/3}_{\mathrm{m}}}{\tau_0},
\label{Htau}
\end{equation}
which gives
\begin{equation}
H(\tau_0)=  \frac{2 \,(1-\Omega^{1/3}_{\mathrm{m}})}{\tau_0 \,(1-\Omega_{\mathrm{m}})}.
\label{Hzero}
\end{equation}

Thus, one also has from (\ref{taz})

\begin{equation}
 \tau(z) = \frac{2 \,\left[\Omega_{\mathrm{m}}+(1-\Omega_{\mathrm{m}})(1+z)^{-3/2}\right]^{1/3} -\Omega^{1/3}_{\mathrm{m}}}
{H_0 \,(1-\Omega_{\mathrm{m}})}
\label{tzo}
\end{equation}

For $H(z)$, we obtain the values from (\ref{h}) and  (\ref{tz}),
\begin{equation}
H(z) = \frac{2 \, \left[\Omega_{\mathrm{m}}+ (1- \Omega_{\mathrm{m}})(1+z)^{-3/2} \right]^{2/3}}
{(1- \Omega_{\mathrm{m}}) (1+z)^{-3/2}},
\label{HZ}
\end{equation}
\begin{equation}
\mathrm{and} \quad H(z) \, = \, H_0 \left[\Omega_{\mathrm{m}}(1+z)^{9/4}+(1-\Omega_{\mathrm{m}})(1+z) ^{3/4} \right]^{2/3}
\label{hhz}
\end{equation}
as first shown by \citet{Jesus18}. There, $H(z)$ is expressed  in the units chosen  for $H_0$.\\

\subsection{The T-  and Density-Evolution, Equilibrium and Recombination}

 The previous developments apply to the  matter-dominant phase of the Universe. For  models with \om ~> 0.05, the recombination occurs  in the matter 
 dominant phase, i.e., with $t^*_{\mathrm{rec}} >  t_{\mathrm{eq}}$, thus there is no need of a treatment appropriate for the radiative phase. 
 For the models with \om ~$\leq 0.05$  the recombination occurs during  the radiative phase. In such case, the appropriate  expansion factor $a(t)$ has 
 been applied in the concerned radiative part. We may note that alike in the matter phase, the radiative phase also possesses 
a simple  analytical   expression  \citep{Maeder19}.

\begin{figure}[H]
\centering
\includegraphics[width=1.05\textwidth]{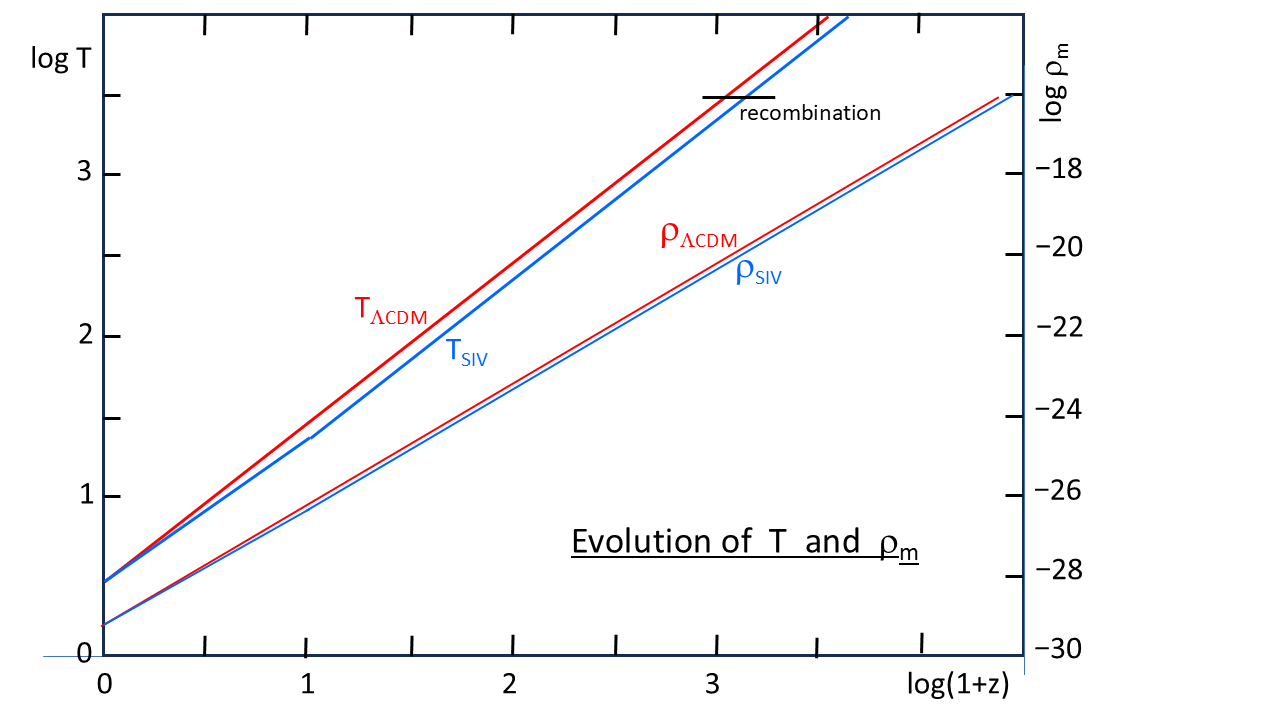}
\caption{Temperature variations in the CMB and of the matter density as a function of redshift $z$ for the  $\Lambda$CDM  and SIV models both with $\Omega_{\mathrm{m}}=0.30$. 
One notices that for SIV the recombination occurs a higher $z$. A mean value of $H_0=70$ \kmsmpc\ is adopted.}
\label{tr}
\end{figure}

According to Equation  (\ref{3w}), the conservation laws are different in the matter  and radiation phase, respectively these are \citep{Maeder17a},
\begin{equation}
\mathrm{matter:}\quad   \varrho_{\mathrm{m}} \, a^3  \, \lambda =const. \quad \quad \mathrm{radiation:} \quad \varrho_{\mathrm{rel}} \, a^4  \, \lambda^2 =const.    
\end{equation}
\noindent
These relations lead to 
\begin{eqnarray}
T \cdot  a(t) \cdot t^{-1/2}=2.726 K, \quad
\varrho_{\mathrm{rel}}\cdot a^4(t) \cdot t^{-2} \;=\; 4.6485 \times 10^{-34} \, \; K_0 \; \mathrm{[g/cm^3]},  \\        
\varrho_{\mathrm{m}} \cdot  a(t)^3  \cdot t^{-1}=1.8788 \times 10^{-29} \, h^2 \Omega_{\mathrm{m}} \; \mathrm{[g/cm^3]}.
\label{conserv}
\end{eqnarray}
\noindent
The term $K_0$ is the ratio of the number of relativistic particles to photons; for a number of neutrinos $N_{\nu}=3$, one has $K_0=1.6813$. 
In the standard $\Lambda$CDM case, the terms in $t$ are absent.
Figure~\ref{tr} compares the backwards evolution of the temperature of the CMB and  matter density $\varrho_{\mathrm{m}}$ 
for the $\Lambda$CDM and SIV models both with \om ~= 0.30, according to their respective conservation laws. The presence of the $t$-terms in SIV makes both the temperature 
and matter density to be lower than in the standard case for a given redshift. A direct consequence is that the recombination
(alike other critical points defined by a critical  temperature $T$)  occurs at
a higher redshift $z$.  For lower \om, the differences between the two curves would
be larger, which is a general rule for  SIV.

Table~\ref{tabparam} gives for different values of \om: in column 2, the initial ratios of the densities of relativistic particles to matter density, the other columns give,  at the equilibrium of matter and radiation,  $a_{\mathrm{eq}}$, $z_{\mathrm{eq}}$, time $t_{\mathrm{eq}}$ (between $t_{\mathrm{in}}$ 
and 1.0) and time $\tau_{\mathrm{eq}}$ in years. We notice
the large  redshifts $z_{\mathrm{eq}}$  at equilibrium which rapidly decrease for lower \om, while the age of the equilibrium, small for
\om = 0.30 rapidly increases for lower \om. Note that the relation between ages $t$ and $\tau$  is given by
$\frac{\tau - \tau_{\mathrm{in}}}{\tau_0 - \tau_{\mathrm{in}}} = \frac{t - t_{\mathrm{in}}}{t_0 - t_{\mathrm{in}}}$
expressing that the age fraction with respect to the present age is the same in both timescales and giving,
\begin{equation}
\tau \,= \, \tau_0 \, \frac{t- \Omega^{1/3}_{\mathrm{m}}}{1- \Omega^{1/3}_{\mathrm{m}}} \,  \quad \mathrm{and} \; \;
  t \,= \, \Omega^{1/3}_{\mathrm{m}} + \frac{\tau}{\tau_0} (1- \Omega^{1/3}_{\mathrm{m}}) \,,
\label{T2}
\end{equation}
with constant derivatives.
For larger  $\Omega_{\mathrm{m}}$,   timescale $t$ is squeezed over a small interval, which  reduces the range of $\lambda$-variations.

\begin{table}[H]

 \caption{Values of parameters at the equilibrium point between matter and radiation in the scale-invariant models with $k=0$ and  $H_0=70$ \kmsmpc for   various $\Omega_{\mathrm{m}}$. The second column gives the ratio of the density of the relativistic particles including the neutrinos to the matter density. The coefficient  $K_0 =1.6813$ applies for a number of neutrinos types, $N_{\nu}= 3$. Time $t_{\mathrm{eq}}$ is the equilibrium time in the dimensionless $t$-scale varying between  $t_{\mathrm{in}}$ and 1.0. Time $\tau_{\mathrm{eq}}$ is the equilibrium time in~years. }

\begin{adjustwidth}{-\extralength}{0cm}
\begin{tabularx}{\fulllength}{CCCCCC}
\toprule
$\boldsymbol{\Omega_{\mathrm{m}}}$  &  $\boldsymbol{\frac{K_0 \varrho_{\gamma,0}}{\Omega_{\mathrm{m}}\varrho_{\mathrm{c,0}}}}$  & $\boldsymbol{a_{\mathrm{eq}}}$ & $\boldsymbol{z_{\mathrm{eq}}}$ 
   & $\boldsymbol{t_{\mathrm{eq}}}$ & $\boldsymbol{\tau_{\mathrm{eq}}}$   \\
\midrule
 0.30& 2.82982878 
 $\times 10^{-4}$ & 1.8943844702 $\times 10^{-4}$ & 5277.759  &0.669434307657 & 42'497.93 \\
 0.20& 4.24474317 $\times 10^{-4}$ & 2.4823538095 $\times 10^{-4}$ & 4027.435  &0.584806597236 & 76'020.31 \\
 0.10& 8.48948634 $\times 10^{-4}$ & 3.9405629697 $\times 10^{-4}$ & 2536.708  &0.464169775550 & 210'389.06 \\
 0.05& 1.69789727 $\times 10^{-3}$ & 6.2557268086 $\times 10^{-4}$ & 1597.535  &0.368439652927 & 598'193.03 \\
 0.02& 4.24474317 $\times 10^{-3}$ & 1.1529368281 $\times 10^{-3}$ &  866.350  &0.271615214841 & 2'464'443.49 
 \\
\bottomrule

\end{tabularx}
\end{adjustwidth}

\label{tabparam}
\end{table}

Table~\ref{tabage} gives the various critical times in the $t$- and $\tau$- units. 
 The initial time $t_{\mathrm{in}}$ is   determined by  Equation (\ref{tinL}) for $\Lambda$CDM and by the solution of the cosmological equations in the radiative phase for SIV.  The recombination time $t^*_{\mathrm{rec}}$ and the redshift  $z^*_{\mathrm{rec}}$ are   generally best determined  by the time and redshift at which the optical depth of  Thomson scattering reaches unity during  the $T$- and $ \varrho$- evolution of the Universe model considered  \citep{Piatella18}. The decrease in the ionized fraction is given by Boltzman and Saha equations. The results currently show that the probability for the last scattering of a photon  is a very peaked function of temperature and redshift.
Account  may also be given to departures from equilibrium for excited hydrogen states, as well as to the fact that all photons do not strictly and simultaneously last-scatter.

The ionized fraction $x_{\mathrm{e}}$= 10\%, 1\%  and 0.1\% occur for energies equal to 0.295, 0.259 and 0.250 eV ($T=$3423 K, 3000 K, 2900 K).  Hu \citep{Hu08}  obtained a typical redshift $z^*_{\mathrm{rec}}=1089$ in the standard models, with limited variations with \om, in agreement with  estimates  around  $z^*_{\mathrm{rec}}=1100$ \citep{Ryden03}. The standard test $\Lambda$CDM models we are performing  numerically also predict a redshift $z =1099.51$  at  a temperature of  3000 K for models  with  $\Omega_{\mathrm{m}} $ in the range  0.10 $<$ \om $<$ 0.30.  Thus,  we retain the value $T^*_{\mathrm{rec}}= 3000$ K, with an ionizing fraction of 1\%,  as  defining the recombination. Figure~\ref{tr}   shows that for $T=$3000 K the matter densities  of $\Lambda$CDM  and SIV  models  differ very little.

\begin{table}[H]  \setlength{\tabcolsep}{5.9mm}{}

 \caption{Different parameters characterizing the recombination. In the third column, the initial times of the origin are  given; these values slightly
 differ  from  $\Omega^{1/3}_{\mathrm{m}}$ (typically by $10^{-6}$), since they have been calculated with the $a(t)$-term appropriate 
 for the radiative phase \citep{Maeder19}. Times $t_{\mathrm{in}}$  and  the recombination time $t^*_{\mathrm{rec}}$ are  expressed in the $t$-scale, and time $\tau^*_{\mathrm{rec}}$  in years. All these times    are  independent of the initial $H_0$ value. 
 All $t$-times have to be given with a lot of decimals, owing to their small variation~intervals.
\label{tabage}} 

\begin{tabular}{cccccrr}
\toprule 
$\boldsymbol{\Omega_{\mathrm{m}}}$  &  $\boldsymbol{t^*_{\mathrm{rec}}}$  &  $\boldsymbol{t_{\mathrm{in}}}$ &$\boldsymbol{z_{\mathrm{rec}}^\star}$  &     $\boldsymbol{\tau^*_{\mathrm{rec}}}$   \\
\midrule
 0.30& 0.669443504837  &  0.669433289476 & 1344.05  &   426'440.4 \\  
 0.20& 0.584817830348  &  0.584804310040 & 1438.08  &   449'379.1 \\    
 0.10& 0.464180331509  &  0.464161606384 & 1614.30  &   482'247.5 \\   
 0.05& 0.368433557640  &  0.368412275291 & 1812.08  &   465'012.9 \\    
 0.02& 0.271507068737  &  0.271485114568 & 2111.05  &   415'870.1 \\
\bottomrule
\end{tabular}

\end{table}

Evidently, we cannot assign the above redshift to define the recombination in  scale-invariant models, since the conservation law  
is different. 
Table~\ref{tabage} shows that  $z^*_{\mathrm{rec}}=1344$ for SIV models with \om ~= 0.30 and $T=$ 3000 K, a result explained by the slightly lower temperatures 
at a given redshift in SIV models as  illustrated in Figure~\ref{tr}.   The  differences of $z_{\mathrm{eq}}$ and $\tau_{\mathrm{eq}}$ between $\Lambda$CDM and SIV models 
are larger for lower $\Omega_{\mathrm{m}}$-values, since in this case  the SIV $a(t)$ relation is more deviating 
from the corresponding $\Lambda$CDM model.  
Moreover,  the exact choice of $T_{\mathrm{rec}}$ has little effect on the Hubble tension, due to the 
parallelism of the relations of $T(z)$ and  $H(z)$   vs. $z$, as shown for high $z$-values in Figures~\ref{tr}  and \ref{Fscale}.

%

\section{Confronting SIV with Observations}
\label{sec:obs}

We now confront the expansion history of the Universe predicted in the SIV theory to two key tests: 1- The Hubble diagram of type Ia supernovae, 2- The relation between $H_0$,   the age of the Universe $\tau_0$ and  \om. Further tests of the cosmological expansion history are being performed and will be the subject of future works to comply with our second requirement in the introduction ``The new model should not do worse than standard $\Lambda$CDM in describing all other cosmological
observations''. 

\subsection{Testing SIV with the Hubble Diagram of SN Ia}
\label{sec:SNe}

Type Ia supernovae are standard candles used in the distance ladder method. They allowed us to measure the accelerated expansion rate of the Universe \citep{Riess1998, Perlmutter1999}. Because type Ia supernovae are part of a ladder, the method is subject to potential errors propagating in the consecutive steps of the ladder. For this reason, they do not stand alone as a measurement method for $H_0$. However, they constrain the density of matter \om well. This is what is performed in the SIV paradigm without trying to measure $H_0$.

\begin{figure}[H]

\includegraphics[width=0.6\textwidth]{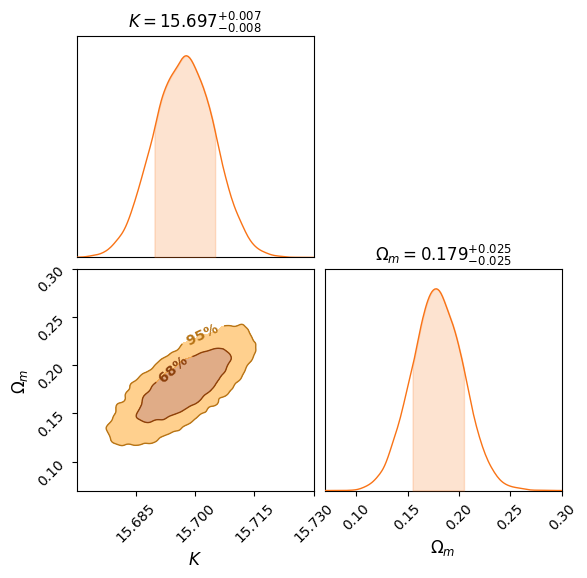}\\
\includegraphics[width=1\textwidth]{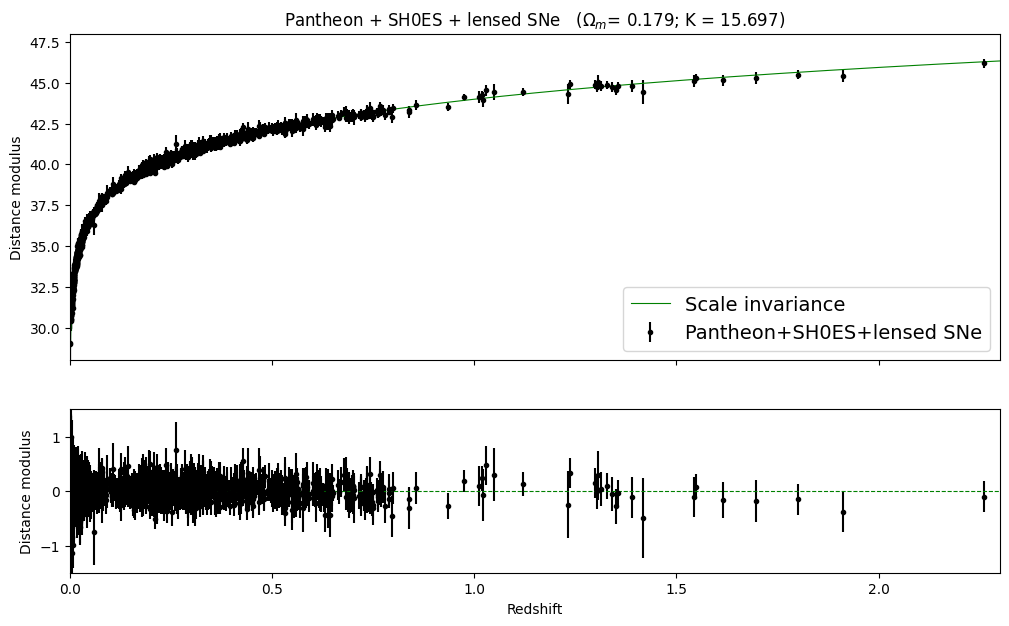}
\caption{Bayesian fit to the supernovae sample compiled by \citep{Pierel2024}. We fit for \om\, and for an arbitrary constant, $K$, on the distance modulus without attempting to constrain $H_0$. We use a flat prior on \om\, such that \om$ \in [0, 0.5]$. The data points are exactly the same as in \citet{Pierel2024}.}
\label{fig:SNe}
\end{figure}

The data set considered is the one compiled by \citet{Pierel2024} that includes Pantheon + SH0ES \citep{Brout2022} and two lensed high-redshift supernovae extending the redshift range to $z=2.9$. Using Equation~(\ref{hhz}), we perform a Bayesian fit to the distance modulus model for SIV, where the free parameters are the matter density, \om, and an arbitrary constant, $K$, in the distance modulus.  The results are shown in Figure~\ref{fig:SNe}, leading to \om$=0.18 \pm 0.03$. We obtain a reduced $\chi^2_{\rm SIV}=0.44$, pointing to overfitting but this likely due to the error bars on the supernovae being over-estimated. For comparison, the best fit in $\Lambda$CDM with \om$= 0.315$, as in \citet{Pierel2024}, gives $\chi^2_{\rm CDM}=0.44$ as well. { We further explore the robustness of our fit. We first artificially normalize the error bars to obtain a $\chi^2_{\rm SIV}\sim 1.0$. This results in the same value of \om. Dividing the error bars by 2 increases the $\chi^2$ values to $\chi^2_{\rm SIV}=1.75$ and $\chi^2_{\rm CDM}=1.77$ but does not change the values of \om. Increasing our prior on \om\ to the range [0, 1] gives the exact same fit as with [0, 0.5] used in Figure~\ref{fig:SNe}, although with slightly slower convergence. Using a prior that does not include the value of \om=0.2 results in hitting the edge of the prior without proper convergence.}

In other words, SIV and $\Lambda$CDM give fits of equal quality, but SIV requires significantly less matter than flat $\Lambda$CDM; however, we recall here that
the \om\ parameters are expressed  in different theoretical models. This is in line with previous observational results in SIV. For example, the growth of structures is easier in the SIV context than in $\Lambda$CDM. This is true for \om~$=0.3$, but values of \om ~ = 0.1--0.2 are even more able to lead to some very early massive structures, as observed with recent JWST observations of high redshift massive galaxies. 

Our previous results with strong gravitational lensing \citep{MaedCourb24} indicated that the Einstein radius of the SLACS lenses were well reproduced for matter densities as low as \om~= 0.05 but were compatible with higher values depending on assumptions of the IMF of the stellar populations in the lens and on the slope of the luminosity function. The same holds true for the very high redshift lens JWST-ER1 by \citep{Mercier24} that is well compatible with values of \om\, in the range 0.10 $<$~ \om ~$<$ 0.20. In other words, strong lensing in SIV is compatible with values as low as \om~$=0.05$ but does not rule out higher values and does not contradict the value we find with type Ia supernovae. In fact, \om$\sim$0.2 is also in line with simulations of the cosmological nucleosynthesis by \citep{GueorMaed25} that indicate values for \om\, that are significantly smaller than in $\Lambda$CDM. The present 
use of $H(z)$ in establishing the theoretical expression of the distance modulus  of supernovae leads to the first direct observational measurement of \om\, in the SIV paradigm. The value of \om~$=0.18 \pm 0.03$ represents a good starting point for further observational tests in SIV. It is also consistent with previous works, but now with error bars.

The value of \om\ obtained in SIV is different from the classical one of \om~= 0.30, and this may appear surprising. But the 
two cosmological models are different: the equations of the  $\Lambda$CDM model contain an additional constant with respect to Friedmann's, while  the SIV models contain an additional term varying with time.
All relations between the different physical variables are different. Thus, different models applied to the same observational data   do not necessarily give the same values of the parameters. For example, the baryon contents of the two types of models are   not necessarily identical \citep{GueorMaed25}. 

\begin{figure}[H]

\includegraphics[width=0.95\textwidth, angle=-0.2]{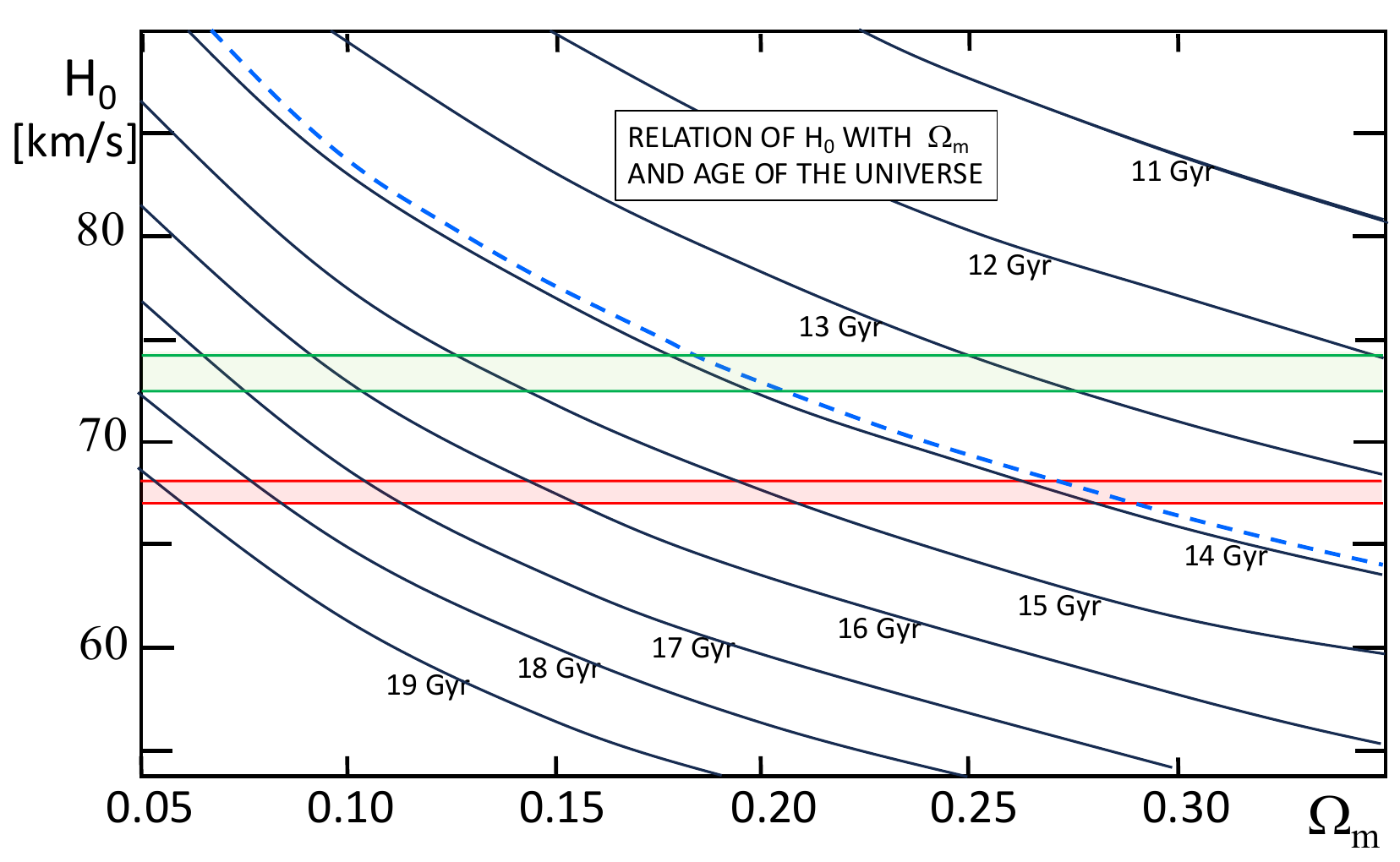}
\caption{Relations predicted in SIV  between $H_0$ and \om\,  for different values of the age of the Universe. The blue dashed line shows the curve for the standard age value of 13.92 $\times 10^9$ years, shown as the blue dashed curve \citep{Valcin25}. The measured local value of $H_0$ and associated 1$\sigma$ error bar is shown in green (SH0ES) and the CMB value from Planck is shown in red, illustrating the tension between the late and early Universe values. For the SIV value of \om$\sim$0.2 measured with type Ia supernovae, and given the estimated age of the Universe, $H_0 \sim 74$ \kmsmpc\, is clearly favored. 
\label{Tinsley}}
\end{figure}

The above remark in fact does not only apply to Figure~\ref{fig:SNe} but also to Figure~\ref{Tinsley} below involving the age of the Universe, $\tau_0$: the relations in the triplets $H_0$-\om-$\tau_0$ studied in the next Subsection are evidently also model-dependent.

As a final note, all the above conclusions are true in the absence of internal evolutionary effects which may affect the properties of SNIa with redshift. This is true both for SIV and $\Lambda$CDM and may be considered in the future, although recent JWST observations indicate no significant luminosity evolution of type Ia supernovae with redshift \citep{Pierel2025}.

\subsection{Relation Between the Age of the Universe and the  $H_0$-$\Omega_{\mathrm{m}}$ Plane}
\label{sec:tinsley}

The comparison of $H_0$, for different values of \om,  with  the age of the Universe, is a fundamental test fully independent of Type Ia supernovae. It was used by Gunn and Tinsley \citep{GunnTinsley75} to propose that the Universe was accelerating and a further study based on the same three parameters by \citet{Tinsley78} confirmed this  result.


Figure~\ref{Tinsley}  shows the relation, in the SIV theory, between $H_0$ and \om\, for different ages of the Universe. The red and green stripes indicate the generally accepted location of the Hubble constant values of $H_0=$74 \kmsmpc\ in the local Universe and $H_0=$67.4 \kmsmpc\ in the early Universe. The blue dashed line gives the relation for the currently accepted age of the Universe of $\tau_0 = 13.92 \times 10^9$ yr \citep{Valcin25}. The age of the Universe is generally based on several approaches, in particular the age of the oldest  globular clusters, the fluctuations of the CMB, or the disintegration of radioactive isotopes. According to \citet{Frieman08}, the estimates based on the ages of the oldest globular clusters suggest values between 12 and 15 Gyr.  

If some of the latest estimates of the age of the Universe are correct, i.e., $\tau_0 = 13.92 \times 10^9$ yr \citep{Valcin25}, indicated as the blue dashed line in Figure~\ref{Tinsley}, then the value of \om ~= 0.18 favored by type Ia supernovae in SIV is in agreement with the value of about 0.20 imposed by the consistency of the triplet $\Omega_{\mathrm{m}}$-$H_0$-$\tau_0$ the age of the Universe. Below, this value of
$\Omega_{\mathrm{m}}$  also appears able to solve the Hubble tension within the SIV context.

Let us precise that the  value of \om ~= 0.20 found here in SIV  is the global value of the matter content. In view of the uncertainties remaining  in the exact value of the baryon content from cosmological nucleosynthesis \cite{GueorMaed25},  the present data do not allow the  determination of
the possible amount dark  matter, nor  if some  is necessary.

\section{The Past Thermal and Dynamical Evolutions  in $\Lambda$CDM and SIV Models and the Possible  Origin of the Hubble Tension}
\label{tension}

As mentioned above, the Hubble tension is the discrepancy  between  the Hubble parameter $H_0=74 \pm 2 $ km s$^{-1}$ Mpc$^{-1}$ determined from supernovae of type Ia  at  redshifts $z$ generally below  2.5, and the value $H_0= 67.4 \pm 0.5$  km s$^{-1}$ Mpc$^{-1}$ obtained from the analysis of  CMB fluctuations. The tension appears as a highly significant cosmological problem  \citep{Verde19, KamionRiess23, H0DN}, which could  possibly find its origin in the physics of the early  Universe and its relation to later phases.

We now consider the problem of the Hubble tension  within  the SIV theory. To do this, we closely compare 
the past thermal and dynamical cosmological  evolutions of $\Lambda$CDM and of SIV  models, in particular we examine the expansion rates $H(z)$ and temperatures
$T$ in the two models. For SIV theory, we recall the rather simple analytical relations which govern the evolution properties above.  For the $\Lambda$CDM model, we apply the  analytical relations developed in Appendix~\ref{LCDM}.

Figure~\ref{Fscale} compares  the $H(z)$ curves up to recombination. The flat $\Lambda$CDM curve in red starts from $H_0 =67.4$ \kmsmpc\, at the present epoch and increases up to $H(z^*_{\mathrm{rec}}) = 1.381 \times 10^6$ \kmsmpc, when the temperature reaches 3000 K at recombination. These values 
(and the curve in the figure) correspond to the standard Planck collaboration results with \om~= 0.315, where the redshift of recombination is   $z^*_{\mathrm{rec}} = 1100$, a value almost independent of \om\  in $\Lambda$CDM models \citep{Hu08}. 
In the case of SIV the redshift of recombination  depends on the matter density since there is a  factor
$t^{-1/2}$ in the time evolution of the temperature $T$ (see Equation~(\ref{conserv})). This leads at high $z$, when the time is close to $t_{\mathrm{in}}$,
to a dependence of  $z^*_{\mathrm{rec}}\propto$ \om$^{-1/6}$. Recombination times and redshifts for different \om\ in SIV are given in Table~\ref{tabage}. For the specific value of \om~= 0.20 in SIV, which is supported by the two major cosmological tests studied above, $H_{\rm SIV}(z^*_{\mathrm{rec}})$ = $H_{\Lambda\rm CDM}(z^*_{\mathrm{rec}})  = (1.381\pm 0.001) \times 10^6$ \kmsmpc\ with $z^*_{\mathrm{rec,\, SIV}}=1438.08 $. In this case, the curve intercept is $H_0 = 74$ \kmsmpc\, matching very well the local measurement of $H_0$.

\begin{figure}[H]

\includegraphics[width=0.9\textwidth]{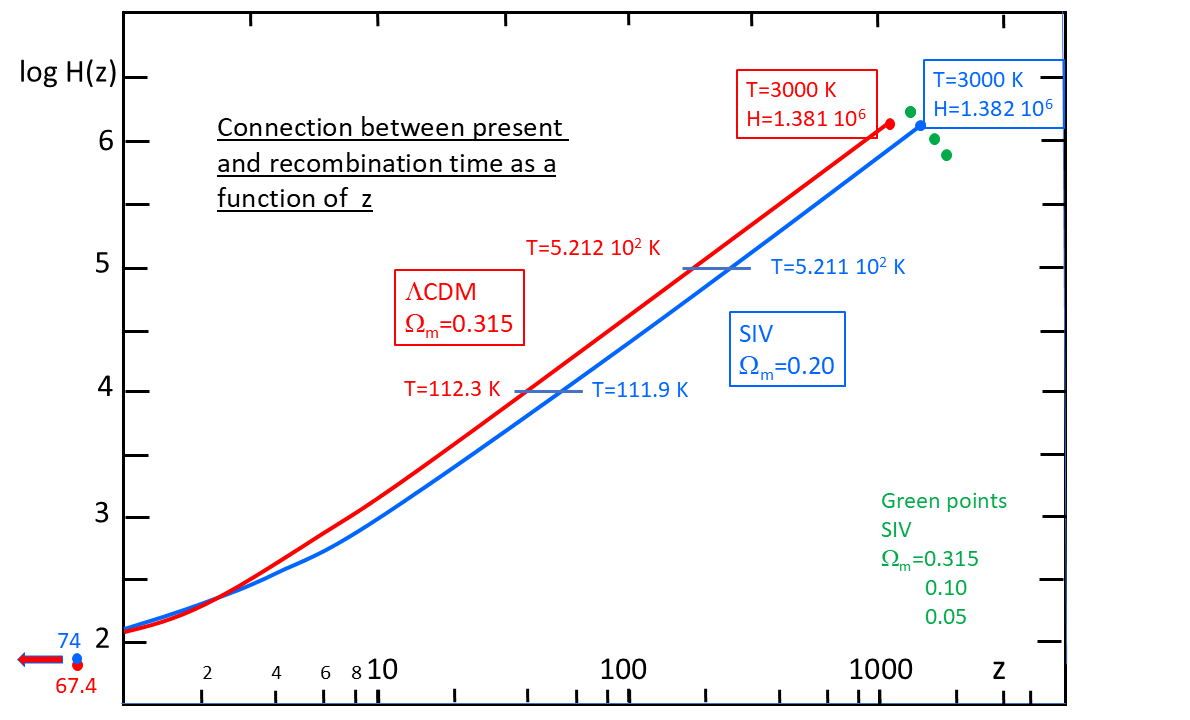}
\caption{Expansion rate of the Universe as a function of redshift in $\Lambda$CDM (red) for \om ~= 0.315 and in SIV (blue) for \om ~= 0.2, as also found with 
supernovae and the $H_0$ - \om - age relation. The $\Lambda$CDM curve for \om ~= 0.315 connects $H_0= 67.4$ \kmsmpc at $z=0$  to $H(z^*_{\mathrm{rec}}) = 1.381 \times 10^6$ \kmsmpc at $z=1099.51$ where \mbox{$T=3000$ K}. The SIV curve for \om =0.20 connects $H_0= 74$ \kmsmpc at $z=0$  to $H(z^*_{\mathrm{rec}}) = 1.382 \times 10^6$ \kmsmpc
at $z=1438.08$ where \mbox{$T=3000$ K}. Thus, identical thermal and dynamical states at recombination are connected to different $H_0$ values by $\Lambda$CDM 
and SIV models for their respective best  observationally supported \om value. The correspondence of thermal and dynamical states is also present over
a large range of high $z$ values as shown at  $H(z) = 10^5$  and $10^4$ \kmsmpc; it progressively disappears at lower $z$. This correspondence only happens for \om = 0.20 in SIV. The points in green give $z^*_{\mathrm{rec}}$ and $H(z^*_{\mathrm{rec}})$ for other values of \om in SIV.  \label{Fscale}}

\end{figure}

In Figure~\ref{Fscale}, we show as green points the expansion rates at recombination in SIV for several other  values of \om\ also  connected to $H_0 = 74$ \kmsmpc\ at present time. The  $H(z^*_{\mathrm{rec}})$ values of these points are rather different from that for \om~= 0.20, as shown in the following. For \om~=~0.315, $H(z^*_{\mathrm{rec}})= 1.66954 \times 10^6$ \kmsmpc, for \om~=~0.10  $H(z^*_{\mathrm{rec}})= 1.03510 \times 10^6$ \kmsmpc, and
for \om=0.05  $H(z^*_{\mathrm{rec}})= 7.75485 \times 10^5 $ \kmsmpc. The differences of $H(z^*_{\mathrm{rec}})$ values are large, although they all correspond to the same temperature of T = 3000~K. Only the SIV model with 
\om=0.20 is matching simultaneously the reference   $H(z^*_{\mathrm{rec}})$ and  the ``local'' value $H_0 = 74$ \kmsmpc.

The similarity of the thermal and dynamical states of the SIV \om~=~0.20 and $\Lambda$CDM \om~=~0.315 cases is  illustrated in Figure~\ref{Fscale}   by two more examples at $H(z) = 10^5$ and $10^4$ \kmsmpc. In the latter case, a small temperature deviation from perfect equality already appears. Two examples of such pairs are indicated in the Figure for T = 521 K and 112 K which correspond respectively to $H(z) = 10^5$ and $10^4$ \kmsmpc. 
For lower redshifts, the difference in the $T(z)-H(z)$  agreement  progressively increases until the present time with $T= 2.726$ K
and  $H_0$  equal to 74 for SIV  and 67.4 \kmsmpc for $\Lambda$CDM. 

The equality of the  dynamical and thermal states, leading eventually to \linebreak $H_0= 67.4$ \kmsmpc in $\Lambda$CDM and 74 \kmsmpc,  covers a large part of the evolution of the Universe. It is noticeable that this  occurs for \om=0.20 in SIV, which is the value found  in Sections~\ref{sec:SNe} and \ref{sec:tinsley} with supernovae and the age-$H_0$-\om\, relation. The reasons for the Hubble tension being solved in SIV are that the shape of $H(z)$ is different from that in $\Lambda$CDM, and  that $z^*_{\mathrm{rec}}$ is anchored at higher redshift yet corresponds to the  same value of $H(z^*_{\mathrm{rec}})$. 

\begin{figure}[H]

\includegraphics[width=0.8\textwidth]{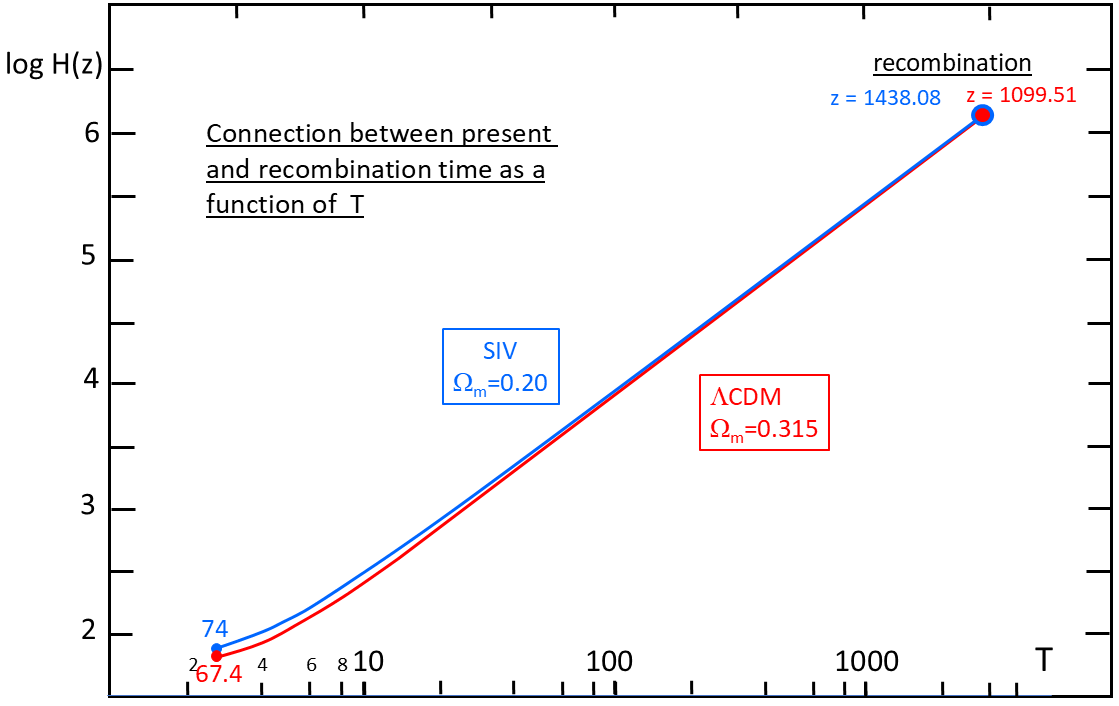}
\caption{{{Expansion} 
} 
 rate of the Universe as a function of temperature, T,  in $\Lambda$CDM (red) for\mbox{ \om ~= 0.315} and in SIV (blue) for \om ~= 0.20, as also found with supernovae 
and   the $H_0$-\om-age relation.  From the same temperature $T$  and same $H(z^*_{\mathrm{rec}})$ at recombination, the
$\Lambda$CDM and SIV models lead respectively to   two different values  $H_0 = 67.4$   and 74 \kmsmpc at $z=0$ for $T=2.726$ K. 
 As indicated, the redshifts at recombination are not the same. However, the figure shows that the models in $\Lambda$CDM and SIV that have the same 
dynamical and thermal states at recombination lead to slightly different $H_0$ values at the present epoch.\color{black} } 
\label{T_CMB}
\end{figure}

Another way of comparing the expansion  histories in $\Lambda$CDM and SIV is to show the evolution  of $H(z)$ with temperature, as in Figure~\ref{T_CMB}. 
At recombination, the two curves coincide, they share the same temperature, T = 3000 K, and expansion rate \linebreak $H(z^*_{\mathrm{rec}}) \simeq 1.381 \times 10^6$ \kmsmpc, while their redshifts are different.
For \om$\neq$0.2, there is no coincidence at recombination.
 At $z=0$ the SIV curve agrees with the ``local''  $H_0=74 $ \kmsmpc, while the  $\Lambda$CDM  model leads to the usual one of $H_0=67.4$ \kmsmpc.

It could be argued (with some reason) that our criterion to define the recombination redshifts  in the two cosmological models is too simple when compared to sophisticated numerical methods using the full optical depth of the models to accurately determine the recombination level. From the  parallelism of the  $T(z)$ relations in Figure~\ref{tr} as well as of the relative parallelism of the $H(z)$ relations at high redshift in Figure~\ref{Fscale}, it is clear that changes of the recombination temperature would not change the results. Moreover, we have seen that the coincidence of the dynamical and thermal states covers a large part of the evolution of the Universe.

As stated in the introduction the new model should not do worse than standard $\Lambda$CDM in describing all other cosmological observations. This is particularly true for the power-spectrum of the CMB, that is very well fitted in $\Lambda$CDM. We provide here preliminary results on the angular position, $\theta$, of the first acoustic peak given by the ratio of the sound horizon, $r_d$ at recombination to the angular diameter distance, $D_{\rm rec}$, to the CMB:

\begin{equation}
\theta = \frac{r_d}{(1+z_{\rm rec}) D_{\rm rec}}.
\label{eq:theta}
\end{equation}

For the flat $\Lambda$CDM model, the sound horizon is $r_d = 144$ Mpc, with \om~=~0.315. The recombination redshift is $z_{\rm rec}=1100$, independent on \om. For $H_0 = 67.4$ \kmsmpc\ this gives an acoustic peak at $\theta_{\rm{CDM}} = 0.594$ degrees. 

For the case of SIV, we find that both the sound horizon and the recombination redshift depend on \om. For the specific value of \om~=~0.2,  we have from Table~\ref{tabage} that $z_{\rm rec}=1438$.   For $H_0 = 74$ \kmsmpc\ we find a sound horizon of $r_d = 168$ Mpc, which translates into $\theta_{\rm{SIV}} = 0.577$ degrees. If the exact value of \om=0.18 found with supernovae is used, the sound horizon is $r_d = 177.5$ Mpc and $z_{\rm rec}=1462$, giving a value of 
$\theta_{\rm{SIV}} = 0.588$ degrees in even better agreement with CMB observations. 

If we now take, in SIV, \om~=~0.3, we have a sound horizon $r_d = 136$ Mpc and $z_{\rm rec}=1344$, translating into $\theta_{\rm{SIV}} = 0.559$ degrees, which is going away from the observed value, consistent with our finding that for such values of \om\ the $\Lambda$CDM and the SIV model do not share the same thermal and dynamical evolutions.

The above considerations are preliminary and will be developed in detail in an upcoming paper, but they  show that the angular scales in the SIV model are already compatible with the observations and that the remaining difference maybe explained by doing more precise calculations and/or by determining \om\ more precisely than with supernovae. Also, note that there is still an uncertainty on the local measurement of $H_0$ that is not accounted for here.

 We finally note that the present results appear to leave room for some amount of dark matter. The  amount of it is still uncertain in view of the remaining uncertainties in the predictions of the SIV nucleosynthesis \citep{GueorMaed25}.

Our analysis, based on the study of the thermal and dynamical properties in the $\Lambda$CDM and SIV models, suggests that the origin of the Hubble tension resides in the way the physical conditions at recombination are  connected by the $\Lambda$CDM model to the present ones in terms of $H(z)$  and temperature evolutions. The account of an additional symmetry, scale invariance, introduces rather minor but measurable effects and appears as an interesting possible solution to the tension.

\section{Conclusions} 
\label{conclusion}

We conclude that if the thermal and dynamical properties of the CMB at \mbox{$T = 3000$ K} are used to infer the present expansion rate, $H_0$,  the value  derived within SIV is  \linebreak $H_0 = 74$ \kmsmpc\, for a specific SIV matter density of \om = 0.2. This last value is supported by the interpretation of the Hubble diagram of type Ia supernovae in the SIV context and independently by the diagram that compares $H_0$, \om\, and the age of the Universe. For that value of \om, the expansion rates of the Universe at $T = 3000$~K in SIV and $\Lambda$CDM coincide almost exactly, implying a similar thermal and dynamical state. Thus, connecting this state to the present time using the $\Lambda$CDM model leads to $H_0$=67.4 \kmsmpc, while the  connection using the SIV model instead gives $H_0$=74 \kmsmpc.

About  the major question raised  in the introduction whether some scale-invariant effects are present in our low density Universe, the present results
suggest a positive answer. For the above mean density, some, in fact very limited, effects remain of what they would be in the empty space. As a consequence, the Hubble tension currently found between $H_0$ values in the early and late Universe may simply be the result of $\Lambda$CDM ignoring the small but still measurable effects of scale invariance.

\authorcontributions{Writing ---original draft A.M.; conceptualization---both authors  A.M. and F.C.; formal analysis---both authors; investigation---both authors; methodology---both authors; validation---both authors; writing---review and editing---both authors. All authors have read and agreed to the published version of the manuscript.}

\funding{This research received no external funding.}

\dataavailability{No new data were created or analyzed in this study.}

\acknowledgments{F.C. would like to thank Martin Millon for help with MCMC fitting, Justin Pierel for providing electronic versions of the supernovae data and Arthur Kosowsky for useful suggestions. A.M. expresses his gratitude to James Lequeux, Georges Meynet, Vesselin Gueorguiev and Gilbert Burki for their support and  encouragement.}

\conflictsofinterest{The authors declare no conflicts of interest. } 

\appendix
\appendixstart
\appendixtitles{yes}

\section{The Analytical $\Lambda$CDM Models} 
\label{LCDM}

Analytical solutions would often be useful, in particular for the resulting functions, such as  time $t$, expansion rate $H$,  temperature $T$, matter and radiation densities,  as a function of redshift $z$. Here, we are considering the case of a flat geometry with $k=0$, which  forms a basis for comparison with the SIV models.

\subsection{Basic  Relations}

Let us start from the basic cosmological equation of the $\Lambda$CDM models,
\begin {equation}
\frac{8\, \pi G \varrho}{3} = \frac{k}{a^2}+ \frac{\dot{a}^2}{a^2}- \frac{\Lambda}{3}.
\label{E1}
\end{equation}

Considering a flat model with $k=0$ and accounting for the conservation law  $\varrho \, a^3= const.$,  we may write,
\begin{equation}
\dot{a}^2  \, a - \frac{\Lambda \, a^3}{3} - C  = 0, \quad \quad \mathrm{with} \quad  C=  \frac{8\, \pi G \varrho \, a^3}{3}.
\label{E2}
\end{equation}

The cosmological constant $\Lambda$ is expressed here as the inverse of the square of a time. With the critical density $\varrho_{\mathrm{c}}= \frac{3 H^2_0}{8\, \pi G }$ and the matter density parameter $\Omega_{\mathrm{m}}= \varrho/ \varrho_{\mathrm{c}}$.  The constants $C$ can be further specified by its value at present  time $t_0$, $C \, = \, \Omega_{\mathrm{m}} \,H^2_0  \, a(t_0)^3$.

We have  from (\ref{E1})

\begin{equation}
   \Omega_{\mathrm{m}}+\Omega_{\Lambda} = 1 ,    \,  \quad \quad \mathrm{with} \quad \Omega_{\Lambda} =\frac{\Lambda}{3 \, H^2},
\end{equation}
where $\Omega_{\mathrm{m}}$, $\Omega_{\Lambda}$,  $\rho_{\mathrm{c}}$ are all varying with time. However, these quantities are usually considered at the present time, unless specified.
We may relate  the constants $\Lambda$  and $C$ thanks to values at present,
\begin{equation}
\Lambda= 3  H^2_0 (1- \Omega_{\mathrm{m}} )=
3  H^2_0 \left(1- \frac{C}{H^2_0 \, a^3(t_0) }\right). 
\end{equation}

We choose the units so that  at present we have $t_0 = 1$ and  $a(t_0)=1$. 

The initial time $t_{\mathrm{in}} $ of the Universe is  defined by $a(t_{\mathrm{in}})=0$.  
The cosmological equation can now be written,
\begin{equation}
\dot{a}^2  - H^2_0 \,\Omega_{\Lambda} \, a^2 -\frac{ H^2_0  \, \Omega_{\mathrm{m}}}{a} = 0.
\label{AE3}
\end{equation}

\subsection{A Time Versus a(t) Solution of the $\Lambda$CDM Models} \label{prem}

The equations of $\Lambda$CDM models are generally solved numerically. Here we are first examining a  useful analytical relation.  From (\ref{AE3}), we get
\begin{equation}
\dot{a} \, = \,\frac{H_0}{a^{1/2} } \, \sqrt{\Omega_{\Lambda} \, a^3+ \Omega_{\mathrm{m}}},
\label{ap}
\end{equation}
%
\begin{equation}
\mathrm{Setting \;}   x= a^{3/2}, \; \mathrm{we \; have} \quad dt \, = \, \frac {2}{3}  \frac{dx}{H_0 \, \sqrt{\Omega_{\Lambda}} \sqrt{\,x^2 + \frac{\Omega_{\mathrm{m}}}{\Omega_{\Lambda}} }}.
\end{equation}
\begin{equation}
\mathrm{Integration\; gives} \quad
t \, = \, \frac {2}{3\, H_0 \,\sqrt{\Omega_{\Lambda}} } \left[\, \ln \left(a^{3/2}+
\sqrt{a^3 + \frac{\Omega_{\mathrm{m}}}{\Omega_{\Lambda}} } \, \,\right) + \tilde{C}\, \right]
\label{Sola}
\end{equation}
where $\tilde{C}$ is a constant. Since $ \frac {2 \, \tilde{C}}{3\, H_0 \,\sqrt{\Omega_{\Lambda}} }$ 
is also a constant  $C'$, we may write  simply
\begin{equation}
t \, = \, \frac {2}{3\, H_0 \,\sqrt{\Omega_{\Lambda}} }  \ln \left(a^{3/2}+
\sqrt{a^3 + \frac{\Omega_{\mathrm{m}}}{\Omega_{\Lambda}} } \, \,\right) +C '\, .
\label{Solb}
\end{equation}

The  condition  $a(t_0)=1$ at $t_0=1$ imposes
\begin{equation}
C' \, = \, 1 - \frac {2}{3\, H_0 \,\sqrt{\Omega_{\Lambda}} } \, \ln \left(1+\sqrt{1 + \frac{\Omega_{\mathrm{m}}}{\Omega_{\Lambda}} } \, \,\right).
\label{ca}
\end{equation}

This establishes a relation between $H_0$ and $C'$ for a given $\Omega_{\mathrm{m}}$. From (\ref{Sola}), we have at the initial time $t_{in}$ when $a(t_{in})=0$,
\begin{equation}
t_{in}= \frac {2}{3\, H_0 \,\sqrt{\Omega_{\Lambda}} } \, \ln \sqrt{\frac{\Omega_{\mathrm{m}}}{\Omega_{\Lambda}}}+ C' 
\end{equation}

If we  choose $t_{in}=0$, we get a second relation between $H_0$ and $C'$. Combining it with (\ref{ca}) gives
\begin{equation}
1= \frac {2}{3\, H_0 \,\sqrt{\Omega_{\Lambda}} } \, \ln \left(1+\sqrt{1 + \frac{\Omega_{\mathrm{m}}}{\Omega_{\Lambda}}}\right)\, 
- \frac {2}{3\, H_0 \,\sqrt{\Omega_{\Lambda}} }\, \ln \sqrt{\frac{\Omega_{\mathrm{m}}}{\Omega_{\Lambda}}}.
\end{equation}

We then get an  expression for  $H_0$ and $C'$,
\begin{equation}
H_0 =  \frac {2}{3\,\sqrt{\Omega_{\Lambda}} } {\ln \left(\frac{1+\sqrt{1 + \frac{\Omega_{\mathrm{m}}}{\Omega_{\Lambda}} } }
{{\sqrt{\frac{\Omega_{\mathrm{m}}}{\Omega_{\Lambda}}}}}\right)}\,  \quad \quad
C' \, = \, - \frac{\ln \sqrt{\frac{\Omega_{\mathrm{m}}}{\Omega_{\Lambda}}}}  
{\ln \left(\frac{1+\sqrt{1 + \frac{\Omega_{\mathrm{m}}}{\Omega_{\Lambda}} } }
{{\sqrt{\frac{\Omega_{\mathrm{m}}}{\Omega_{\Lambda}}}}}\right)}
\label{Ho}
\end{equation}

As an example for a cosmological model with $\Omega_{\mathrm{m}}=0.3$ and $\Omega_{\Lambda}=0.7$, we get
\begin{equation}
H_0 \, =\, 0.9640994\, \quad \quad \mathrm{and} \quad  C'\, =\,0.3501419\,.
\label{f0}
\end{equation}

Apart from the  choice of the  conditions  $a=1$ at $t=1$ at  present, we are free to choose one among the three values  $t_{in}$, $C'$, $H_0$.  Above, we have taken $t_{in}=0$, but we could have chosen  $H_0$ or $C'$ and this would have determined another couple of values. Taking $H_0$ similar to that of another kind of model may sometimes be useful, as permitting  direct comparison of their expansion rates $a(t)$  and 
$H(z)/H_0$ evolution; see  Figure 2  by  \citet{Maeder17a}, where  $\Lambda$CDM  and SIV solutions for $a(t)$ are compared.

\subsection{The General  Analytical $\Lambda$CDM Solution} 
\label{deuss}

It is necessary to have a relation expressing the expansion factor $a(t)$  as a function of time.  We use an alternative  expression of the above solution (\ref{Solb}) in terms of $\mathrm{argsinh}$ \citep{Bronstein74},
\begin{equation}
t \, = \,  \frac {2}{3\, H_0 \,\sqrt{\Omega_{\Lambda}} } \, \mathrm{argsinh} \frac{a^{3/2} \,
 \sqrt{\Omega_{\Lambda}}}{\sqrt{\Omega_{\mathrm{m}}}}+C_1.
\label{Sol2}
\end{equation}

Constant $C_1$ is  different from the previous constants $\tilde{C}$  and $C'$. 
The inversion of  argsinh provides the expansion factor $a(t)$ as a function of $t$,
\begin{equation}
\frac{a^{3/2} \, \sqrt{\Omega_{\Lambda}}}{\sqrt{\Omega_{\mathrm{m}}}}= \sinh \frac{3}{2} \, H_0 \,\sqrt{\Omega_{\Lambda}} \;  (t-C_1) ,
\label{f1}
\end{equation}
\begin{equation}
\mathrm{thus,} \quad  \quad a =   \left(\frac{\Omega_{\mathrm{m}}}{\Omega_{\Lambda}}\right)^{1/3} \sinh^{2/3} \,\left [{\frac{3}{2}  H_0 \sqrt{\Omega_{\Lambda}}\;(t-C_1)}\right] 
\label{sol}
\end{equation}
\begin{equation}
\mathrm{or} \quad a =   \left(\frac{\Omega_{\Lambda}}{\Omega_{\mathrm{m}}}\right)^{1/3} \left(\frac{e^{\frac{3}{2}  H_0 \sqrt{\Omega_{\Lambda}}(t-C_1)}
- e^{- \frac{3}{2}  H_0 \sqrt{\Omega_{\Lambda}}(t-C_1)}}{2} \right)^{2/3}.
\label{Solee}
\end{equation}

These are two equivalent forms of the expansion factor $a(t)$ as a function of time.
The solutions for a given $\Omega_{\mathrm{m}}$ depends on $C_1$ and  $H_0$, which may be constrained by the limiting conditions.
Condition $a=1$ at $t=1$ in (\ref{f1})  is leading to
%
%
%
\begin{equation}
H_0 \, = \, \frac{2}{3 \,(1-C_1) \sqrt{\Omega_{\Lambda}}}\, \mathrm{argsinh} \sqrt{\frac{\Omega_{\Lambda}}{\Omega_{\mathrm{m}}}}.
\label{f2}
\end{equation}

Relation   (\ref{f1})  at the origin $a(t)=0$ with  $t_{in}=0$ is leading to 
$\sinh \,\left(- \frac{3}{2}\,  H_0 \, \sqrt{\Omega_{\Lambda}} \, C_1 \right )=0.$
The argument is also zero and thus $C_1=0$, and we finally have for the solution of the $\Lambda$CDM models,
\begin{equation}
a =   \left(\frac{\Omega_{\mathrm{m}}}{\Omega_{\Lambda}}\right)^{1/3} \sinh^{2/3} \,\left [{\frac{3}{2}  H_0 \sqrt{\Omega_{\Lambda}} \, \, \; t}\right]\, ,
\label{solfin}
\end{equation}
%
%
%
and the present expansion rate  $H_0$  becomes
\begin{equation}
H_0 \, = \, \frac{2}{3 \,\sqrt{\Omega_{\Lambda}}}\, \mathrm{argsinh} \sqrt{\frac{\Omega_{\Lambda}}{\Omega_{\mathrm{m}}}}.
\label{f3}
\end{equation}

As an example, for $\Omega_{\mathrm{m}}=0.3$ and $\Omega_{\Lambda}=0.7$, we get $H_0 \, =\, 0.9640994$, which is  in agreement with  (\ref{f0}). 
For  $\Omega_{\mathrm{m}}=0.2$, $H_0= 1.07602236$ and for $\Omega_{\mathrm{m}}=0.1$, $H_0=1.27787391  $. These values are 
specific to the cosmological models with $a(t_0) =1$, $t_0=1$ and $t_{in}=0$.

As in Appendix~\ref{prem},  
  instead of fixing  $t_{in}=0$, we could have  fixed  the value of $H_0$  (for example for comparison of different kinds of models with observations having an already fixed $H_0$). In this case, the initial time $t_{in}$ is no longer zero, but according to (\ref{sol}) the initial time $t_{in}= C_1$ as obtained from (\ref{f2}),
\begin{equation}
t_{in}= C_1= 1 - \frac{2}{3 \,H_0 \, \sqrt{\Omega_{\Lambda}}}\, \mathrm{argsinh} \sqrt{\frac{\Omega_{\Lambda}}{\Omega_{\mathrm{m}}}}.
\label{tinL}
\end{equation}

Checks have be performed on the exact correspondence between the results of Equation (\ref{sol}) and those  of numerical  $\Lambda$CDM models.

Time $t$ is running from $t_{in}$ at the Big-Bang to 1 at present. Converting the time unit   $t$  into the current units  $\tau$   in Gyr or seconds  or vice versa is straightforward, by expressing that the age fraction with respect to the present age is the same in both timescales. This gives
\begin{equation}
\tau \,= \, \tau_0 \, \frac{t- t_{in}}{1- t_{in}} \, , \quad 
  t \,= \, t_{in}+ \frac{\tau}{\tau_0} (1- t_{in}) \, \; \;\mathrm{and} \; \;
  \frac{d\tau}{dt}  =  \frac{\tau_0}{1- t_{in}}\, .
\label{AT2}
\end{equation}
%

\subsection{The Expansion Rates  $H(z)$}

We need to express the  expansion rate $H=\dot{a}/a$ as a function of redshifts $z$.  From Equation~(\ref{ap}), we have
\begin{equation}
\frac{\dot{a}}{a} \, = \,\frac{H_0}{a^{3/2} } \, \sqrt{\Omega_{\Lambda} \, a^3+ \Omega_{\mathrm{m}}},
\label{apa}
\end{equation}
which gives  with  $\frac{a_0}{a} = 1 + z$,
\begin{equation}
H(z)=  H_0 \, \sqrt{\Omega_{\Lambda}+ \frac{\Omega_{\mathrm{m}}}{a^3}}= H_0 \, \sqrt{\Omega_{\Lambda}+ \Omega_{\mathrm{m}}\, (1+z)^3}.
\label{hz}
\end{equation}

For  model calculations of  $a(t)$  with  $a(t_0) =1$ with $t_0=1$, the expression of $H_0$ given by (\ref{f3}) is quite appropriate.
For  $H(z)$ in km s$^{-1}$ Mpc$^{-1}$,  we must   use  the appropriate  value of $H_0$ in these units. 
Expression (\ref{hz}) is quite convenient,  we could also use the derivative $\dot{a}$ of (\ref{sol}):
$\dot{a} =  \left(\frac{\Omega_{\mathrm{m}}}{\Omega_{\Lambda}}\right)^{1/3}
 \sinh^{-1/3} \,\left [{\frac{3}{2}  H_0 \sqrt{\Omega_{\Lambda}}\; t}\right] \, H_0 \, \sqrt{\Omega_{\Lambda}},$
leading to  $H$ as a function of time,
\begin{equation}
H= \frac{\dot{a}}{a} = H_0 \, \sqrt{\Omega_{\Lambda}}  \, \coth \left [{\frac{3}{2}  H_0 \sqrt{\Omega_{\Lambda}}\; t}\right] .
\end{equation}

It shows that for  $t \rightarrow 0$, the corresponding  $H$-value tends towards infinity.
A relation between $t$ and $z$ is often useful, from (\ref{solfin}) we get
\begin{equation}
t \, = \, \frac{2}{3 \, H_0 \sqrt{\Omega_{\lambda}}} \mathrm{argsinh} \;\left[ (1+z) ^{-3/2} \sqrt{\frac{\Omega_{\Lambda}}{\Omega_{\mathrm{m}}}}\right].
\label{tzz}
\end{equation}
%

\begin{adjustwidth}{-\extralength}{0cm}

\reftitle{References}




\PublishersNote{}
\end{adjustwidth}
\end{document}